\documentclass[nofootinbib,notitlepage,superscriptaddress,10pt,aps,pra,twocolumn]{revtex4-1}

\usepackage[utf8x]{inputenc}
\usepackage{amsmath}
\usepackage{amssymb}
\usepackage{bbm}
\usepackage{graphicx}

\begin{document}

\title{Deformed Lorentz symmetry and relative locality in a curved/expanding
spacetime}

\author{Giovanni AMELINO-CAMELIA}
\affiliation{Dipartimento di Fisica, Universit\`a di Roma ``La Sapienza", P.le A. Moro 2, 00185 Roma, Italy}
\affiliation{INFN, Sez.~Roma1, P.le A. Moro 2, 00185 Roma, Italy}

\author{Antonino MARCIAN\`O}
\affiliation{Department of Physics, The Koshland Integrated Natural Science Center, Haverford College, Haverford, PA 19041 USA}
\affiliation{Department of Physics, Princeton University, New Jersey 08544, USA}

\author{Marco MATASSA}
\affiliation{SISSA, Via Bonomea 265, I-34136 Trieste, Italy}

\author{Giacomo ROSATI}
\affiliation{Dipartimento di Fisica, Universit\`a di Roma ``La Sapienza", P.le A. Moro 2, 00185 Roma, Italy}
\affiliation{INFN, Sez.~Roma1, P.le A. Moro 2, 00185 Roma, Italy}

\begin{abstract}
\noindent The interest of part of the quantum-gravity community in the possibility
of Planck-scale-deformed Lorentz symmetry is also fueled by the opportunities for testing
the relevant scenarios with analyses, from a signal-propagation perspective,
of observations of bursts of particles from cosmological distances.
In this respect the fact that so far the implications of deformed
Lorentz symmetry have been investigated only for flat (Minkowskian) spacetimes represents a very significant
limitation, since for propagation over cosmological distances the curvature/expansion of
spacetime is evidently tangible. We here provide a significant step toward filling
this  gap by exhibiting an explicit example of Planck-scale-deformed
relativistic symmetries of a spacetime with constant rate of expansion (deSitterian).
Technically we obtain the first ever example of a relativistic theory of worldlines of particles
with 3 nontrivial relativistic invariants: a large speed scale (``speed-of-light scale"),
a large distance scale (inverse of the ``expansion-rate scale'),
and a large momentum scale (``Planck scale").
We address some of the challenges that had obstructed success for previous attempts
by exploiting the recent understanding of the connection between deformed Lorentz symmetry and
relativity of spacetime locality.
We also offer a preliminary analysis of the differences between the scenario we here propose
and the most studied
scenario for broken (rather than deformed) Lorentz symmetry in expanding spacetimes.
\end{abstract}

\maketitle

\section{Introduction}
Technically the interest attracted over the last decade by the possibility
of Planck-scale-deformed relativistic kinematics, as conceived within
the proposal ``DSR" (doubly-special, or, for some authors, deformed-special
relativity) introduced in Refs.~\cite{gacdsr1a,gacdsr1b},
was mainly linked to the following observations:

\begin{itemize}

\item it affords us the luxury of introducing as observer-indepedent laws
some of the properties for the Planck scale
that have been most popular in the quantum-gravity literature, such as
a role for the Planck length (the inverse of the Planck scale)
as the minimum allowed value for wavelengths~\cite{gacdsr1a,gacdsr1b};

\item it reflects the content of results rigorously established for 3D quantum gravity
(see, {\it e.g.}, Refs.~\cite{kodadsr,dsr3dSMOLINFREIDKOWA,dsr3dFREIDLIVINE});

\item it reflects the content of results rigorously established for some
4D noncommutative spacetimes
(see, {\it e.g.}, Refs.~\cite{gacmaj,kowaDSRnoncomm});

\item it fits the indications emerging from some compelling semiheuristic arguments
based on 4D
Loop Quantum Gravity~\cite{kodadsr,dsrLQGa}.

\end{itemize}

A comparable ``motivational list" can be claimed by other scenarios for
the fate of relativistic symmetries at the Planck scale,
but in the case of these DSR scenarios a crucial additional motivation
comes from the opportunities in phenomenology.
Some of the novel effects that can be accommodated within a DSR-relativistic
kinematics
can be tested through observations
of bursts of particles from cosmological distances~\cite{dsrphen,unoEdue}.
The key for these analyses are the implications of DSR deformations for
the propagation of signals, and the sought Planck-scale sensitivity is reached
thanks to the huge amplification afforded by the cosmological distances.

Our analysis takes off from the realization that this most notable ``selling point"
for DSR research, based on its phenomenological prospects, has not yet
been made truly accessible by the DSR literature developed so far.
DSR-deformed
relativistic frameworks  have been investigated
so far only for flat (Minkowskian) spacetimes, and this is not the case
relevant for the analysis of signals received from sources at cosmological
distances, for which the curvature/expansion of
spacetime is very tangible. We here provide a significant step toward filling
this  gap by exhibiting an explicit example of Planck-scale-deformed
relativistic symmetries of a spacetime with constant rate of expansion (deSitterian).
Evidently this will lead us to introduce
the first ever example\footnote{Studies such as those in Refs.~\cite{leeTRIPLY,mignemiSITTER,gacQDESITTER}
did contemplate the possibility of 3 invariants, but did not go as far as giving a
consistent relativistic picture
of worldlines of particles.}
 of a relativistic theory of worldlines of particles
with 3 nontrivial relativistic invariants: a large speed scale (``speed-of-light scale"),
a large distance scale (inverse of the ``expansion-rate scale'),
and a large momentum scale (``Planck scale").

There had been previous attempts of investigating the interplay
between DSR-type deformation scales and spacetime expansion
(see, {\it e.g.}, Refs.~\cite{mignemiSITTER,gacQDESITTER}),
but without ever producing a fully satisfactory picture of how the
worldlines of particles should be formalized and interpreted.
In retrospect we can now see that these previous difficulties
were due to the fact that the notion of relative locality had not yet
been understood, and without that notion the interplay between DSR-deformation
scale and expansion-rate scale remains unintelligible.

Relative locality (on which we shall of course return in more detail in later parts
of this manuscript) is the spacetime counterpart of the DSR-deformation scale $\ell$
just in the same sense that relative simultaneity is the
spacetime counterpart of the special-relativistic scale $c$
(scale of deformation of Galilean Relativity into Special Relativity).
This was understood only very recently, in studies such as
the ones in Refs.~\cite{whataboutbob,leeINERTIALLIMIT,principle}. Awareness
of the possibility of relative locality is already very important in making sense
of the implications of DSR-deformations in a flat/non-expanding spacetime.
And, as we shall here show, it plays an even more crucial role in the consistency
of the spacetime picture emerging from the interplay between
DSR-deformation
scale and expansion-rate scale.

While our original (and still here primary) motivation was to address
these fascinating, but merely technical, issues concerning the interplay between
DSR-deformation
scale and expansion-rate scale,
as shown in later parts of this manuscript our results
appear to carry rather strong significance for the phenomenology.
In particular, our constructive analysis leads to a description of
the dependence of the novel effects on redshift that is not  like
anything imagined in the previous DSR literature (relying then
only on heuristic arguments for the
interplay between
DSR-deformation
scale and expansion-rate scale).

We work at leading order in the DSR-deformation scale $\ell$, since (assuming it is of the order of the Planck length) that is the only realistic target for DSR phenomenology over the next few decades. And in order to keep things simple, without renouncing to any of the most significant  conceptual hurdles, we opt to work in a 2D spacetime (one time and one spatial dimension).

The choices of notation we adopt are relatively standard and self-explanatory with the exception of the fact that occasionally we denote with $X$ the set of spacetime coordinates, i.e. $X \equiv (t,x)$, and we denote with $G$ the set of symmetry generators, {\it i.e.} $G \equiv (E,p,{\cal N})$.

\section{Preliminaries on deformed Lorentz symmetry without expansion}\label{secflat}
For our purposes the main starting point is provided by the analysis in Ref.~\cite{whataboutbob}
which provided, assuming a Minkowskian spacetime (no expansion), a framework for
DSR-deformed relativistic theories of worldlines of particles
with two nontrivial relativistic invariants: a large speed scale $c$ (the ``speed-of-light scale", here
mute because of the conventional choice of units $c=1$)
 and a large momentum scale $\ell^{-1}$ (assumed to be roughly of the order of the ``Planck scale").
Ref.~\cite{whataboutbob} however has a limitation that for our purposes is significant:
the analysis in Ref.~\cite{whataboutbob} took off from an ansatz, with a single parameter $\ell$,
for the energy(/momentum) dependence of travel times of massless particles from a given
emitter to a given detector, so it did not fully explore the possible issue of
a dependence of such travel times on the choice of DSR-deformation of the
on-shell relation. Taking as starting point the special-relativistic on-shell
relation, $m^2= E^2 - p^2$, one could contemplate, in particular, adding a term of form $E p^2$
and/or adding a term of form $E^3$. This issue was not much of interest
in Ref.~\cite{whataboutbob} since it is easy to see that
the difference between $E p^2$ deformation
and $E^3$ deformation, if analyzed in a flat/non-expanding spacetime, carries very little significance
(it should be inevitably insignificant at least for massless particles).
But in our analysis of a first case with
spacetime expansion we shall find that there are some significant
difference between $E p^2$ deformations
and $E^3$ deformations.

So in this section we find appropriate to offer a minor generalization and reformulation
of the results reported in Ref.~\cite{whataboutbob}, particularly suitable for our generalization
to the case of an expanding spacetime, most notably for what concerns the comparison
between  $E p^2$ deformation
and $E^3$ deformation.

We start by assuming that, in a Minkowskian spacetime,
the relativistic symmetries leave invariant the following combination
of the energy $E$ and momentum $p$ of particles:
\begin{equation}
{\cal C}_{\alpha , \beta} = E^2 - p^2 + \ell \left( \alpha E^3 + \beta E p^2 \right)\ ,
\label{casimirFLAT}
\end{equation}
where $\alpha$ and $\beta$ are two numerical parameters.

And we exhibit  a corresponding DSR-deformed 1+1D Poincar\'e
algebra of charges compatible with the invariance of ${\cal C}_{\alpha , \beta}$,
describable in terms of the following Poisson brackets:
\begin{gather}
\left\{ E , p \right\} = 0 \ , \qquad \left\{ E , {\cal N} \right\} = p - \ell  \left( \alpha + \beta \right) p  E \ , \nonumber \\
\left\{ p , {\cal N} \right\} = E + \frac{1}{2} \ell  \left( \alpha E^2 + \beta p^2\right)\ .
\label{algebraFLAT}
\end{gather}

A conveniently intuitive picture of the deformation we are studying is obtained
by giving a representation of these symmetry generators
in terms of  time and space coordinates $t,x$
and variables $\Omega,\Pi$ canonically conjugate to them\footnote{Some authors
(see, {\it e.g.}, Ref.~\cite{jurekPHASE,lukiePHASE})
have studied similar
deformations of Lorentz symmetry adopting spacetime coordinates with Poisson brackets $\{ {\tilde x},{\tilde t} \}= \ell {\tilde x}$.
This was done in relation to the fact that there are quantum-spacetime pictures that can be analyzed in
relation to such deformed-Lorentz-symmetry scenarios, and the most notable case is ``$\kappa$-Minkowski"
(see, {\it e.g.}, Ref.~\cite{majrue,lukieANNALS}), a non-commutative spacetime
with $ [ {\hat x}, {\hat t} ]= i \ell {\hat x}$.
We opt for standard Poisson brackets since it is known that the results of analyses of travel times
give the same results both assuming $\{ {\tilde x}, {\tilde t} \}= \ell {\tilde x}$ and assuming $\{ x,t \}=0$.
This was shown in Refs.~\cite{kappabob,anatomy}, and is a simple consequence of the fact that
one can obtain coordinates $\{ {\tilde x}, {\tilde t} \}= \ell {\tilde x}$, from our coordinates
($\{ x,t \}=0$) by posing ${\tilde t} = t$, ${\tilde x} = x + \ell \Omega x$.
(The point is~\cite{kappabob,anatomy} that $x$ and $x + \ell \Omega x$ coincide in the origin,
so their difference can never affect the determination of when a particle reaches (or is emitted) from the
origin of the spacetime coordinates of an observer.
Ref.~\cite{jacktesi} will show explicitly that our analysis (also for the case with spacetime expansion)
redone using coordinates such that $\{ {\tilde x},{\tilde t} \}= \ell {\tilde x}$  leads to the same results
for travel times.}:
\begin{gather}
\left\{ \Omega,t\right\} =1\ ,\qquad\left\{ \Omega,x\right\} =0\ ,\nonumber \\
\left\{ \Pi,t\right\} =0\ ,\qquad\left\{ \Pi,x\right\} =1\ ,\nonumber \\
\left\{ t,x\right\} = \left\{ \Omega ,\Pi \right\} =0\ ,
\label{phasespaceCANONICALflat}
\end{gather}
 We find as representation the following:
\begin{gather}
E = \Omega + \frac{1}{2}\ell \left( (1 - \beta ) \Pi^2 - \alpha \Omega^2 \right) \ , \label{chargesFLATe}\\
p = \Pi \ , \label{chargesFLATp} \\
{\cal N} = t p + x E + \ell \left( \frac{1}{2} \alpha x E^2 - t p E + \frac{1}{2} \beta x p^2\right) \ .
\label{chargesFLATn}
\end{gather}

As done in Ref.~\cite{whataboutbob}, we shall derive the worldlines adopting a covariant
formulation of classical relativistic mechanics, also introducing
an auxiliary affine parameter labeling points on the worldline
of a particle of mass $m$. As standard in the covariant
formulation of classical relativistic mechanics,
evolution is coded in a pure-constraint Hamiltonian; specifically,
the evolution in the auxiliary parameter on the worldline is governed by a Hamiltonian
constraint: ${\cal C}_{\alpha , \beta} - m^2 = 0$.
This straightforwardly leads~\cite{whataboutbob} to the following worldlines
\begin{equation}
\begin{split}
 x_{m,p} & = x_{0} - \left( \frac{p}{\sqrt{p^{2} + m^{2} }} + \ell p \right) (t-t_{0}) ~,
\end{split}
\label{flatworldline}
\end{equation}
which in the particularly interesting case of massless particles reduces to
\begin{equation}
x_{m=0,p}\left(t\right)=x_{0}-\frac{p}{|p|}\left(t-t_{0}\right)\left(1-\ell|p|\right) ~.
\end{equation}
Notice that with our choices of conventions the velocity is positive when $p$ is negative,
and therefore, restricting our focus on $p<0$, one has
\begin{equation}
x_{m=0,p<0}\left(t\right)=x_{0}+\left(t-t_{0}\right)\left(1-\ell|p|\right)\ .\label{worldlineFLAT}
\end{equation}
And let us also stress that evidently the charges $E,p,{\cal N}$ are conserved along the motion,
in the sense that $ \dot{G} = \left\{ {\cal C}_{\alpha , \beta} , G \right\} = 0$,
and they generate respectively deformed time translations, spatial translations, and boosts, by Poisson brackets.
By construction we have ensured that these transformations all are relativistic symmetries of the
theory, as one can explicitly verify~\cite{whataboutbob} by acting
with them on the worldlines (\ref{flatworldline}), finding that the worldlines are covariant.

At this stage of the analysis the physical content of these worldlines is still hidden
behind the relativity of locality. A warning that this might be the case is seen in the
fact that we are analyzing a case where the on-shell relation, in light of (\ref{casimirFLAT}),
is $\alpha$ and $\beta$ dependent,\footnote{Note that the quantity
we denote
with $m$ (and we loosely refer to as the mass) is not the rest energy
when $\alpha \neq 0$.
The rest energy evidently is $\mu$ such that $m^2 = \mu^2 +  \alpha \ell \mu^3$.
The interested readers can easily verify that none of our most significant
observations (most evidently since we mainly focus on
massless particles) depends on the difference between $m^2$ and $\mu^2$.}
$$m^2 = E^2 - p^2 + \ell \left( \alpha E^3 + \beta E p^2 \right)~,$$
whereas our worldlines are independent of $\alpha$ and $\beta$.
Most importantly the {\underline{coordinate velocity}} one infers from those wordlines is
independent of $\alpha$ and $\beta$.
But one of the main known manifestations of relative locality is a
mismatch~\cite{whataboutbob,kappabob,leelaurentGRB} between coordinate velocity\footnote{Examples
of velocity artifacts were
of course known well before the understanding of relative locality for theories with deformed Lorentz symmetry:
for example in de Sitter spacetime (and any expanding spacetime) the coordinate velocity of a particle {\underline{distant}}
 from
the observer can be described by the observer as a velocity greater than the speed of light, even though
in classical de Sitter spacetime the
physical velocity measured by an observer {\underline{close}}
to the particle is of course always no greater than the
speed of light. Since these previously known velocity artifacts are connected with spacetime curvature,
 the recent realization that some of our currently investigated
theories formulating (one form or another of) ``deformed Minkowski spacetime"
are subject to relative locality, with some associated
velocity artifacts~\cite{whataboutbob,kappabob,leelaurentGRB},
was largely unexpected.}
and {\underline{physical velocity}} of particles.

For what concerns specifically the analysis so far reported in this section,
the main challenge resides in the fact that we are used to read velocities off the formulas for
worldlines, but this implicitly assumes that translation transformations are trivial.
Essentially we take the worldline written by a certain observer Alice to describe both the emission
of a particle ``at Alice" (in Alice's origin) and the detection of the particle far away from Alice.
The observer/detector Bob that actually detects the particle, since he is distant from Alice,
should be properly described by acting with a corresponding translation on Alice's worldline.
And the determination of the ``arrival time at Bob"
(crucial for determining the physical velocity~\cite{kappabob})
should be based on Bob's description of the worldline, just as much as the ``emission time at Alice"
should be based on Alice's description of the worldline. When translations are trivial (translation generators
conjugate to the spacetime coordinates) we can go by without worrying about this more careful level of discussion,
since the naive argument based solely on Alice's worldline gives the same result as the more careful analysis
using Alice's worldline for the emission and Bob's description of that same worldline for the detection.
But when translations are nontrivial, and one has associated features of relativity of locality,
this luxury is lost.

One way to have ``relative locality" is indeed the case here of interest, with the
non-trivial translation generators of (\ref{chargesFLATe})-(\ref{chargesFLATp})
acting on spacetime coordinates as follows:
\begin{gather}
\{ E , t \} = 1 - \ell \alpha E \ , \qquad \{ E , x \} = \ell (1 - \beta ) p \ , \nonumber \\
\{ p , t \} = 0 , \qquad \{ p , x \} = 1 \,.
\end{gather}
So, following Ref.~\cite{kappabob}, let us probe the difference between coordinate velocity
and physical velocity through the simple exercise of considering the simultaneous emission ``at Alice"
of two massless particles, one ``soft" (with momentum $p_s$ small enough that $\ell$-deformed terms in formulas
fall below the experimental sensitivity available)
and one ``hard" (with momentum $p_h$ big enough that at least the leading $\ell$-deformed terms in formulas
fall within the experimental sensitivity available).

We of course describe the relationship between the coordinates of two distant observers in relative rest
in terms of the
 Poisson-bracket action of
the translation generators $E$, $p$,
{\it i.e.} $\mathbbm{1} - a_t \{ E , \cdot \} - a_x \{ p , \cdot \}$,
with $a_t$, $a_x$ the translation parameters along $t$ and $x$ axes.
In the specific case in which Bob detects the soft massless particle in his origin,
which restricts us to the possibilities  $a_t = a_x = L$ ($L$ being the spatial distance between Alice and Bob),
 one finds that Bob's coordinates are related to Alice's as follows:
\begin{gather}
t_0^B = t_0^A - L  + \ell L \alpha E \ , \label{jocbobaliceT} \\
x_0^B = x_0^A - L - \ell L (1 - \beta ) p  ~ .
\label{jocbobaliceX}
\end{gather}
The case we are considering, with a soft and a hard massless particle simultaneously emitted toward Bob
in Alice's origin, is such that the two particles are described by Alice in terms
of the worldlines
\begin{gather}
x_{m=0,p_s<0}^A (t^A)= t^A ~, \nonumber \\
x_{m=0,p_h<0}^A (t^A)= t^A\left(1-\ell|p_h|\right) ~.\nonumber
\end{gather}
And these worldlines, in light of (\ref{jocbobaliceT})-(\ref{jocbobaliceX}),
are described by Bob as follows:
\begin{gather}
x_{m=0,p_s<0}^B (t^B)= t^B \ , \nonumber \\
x_{m=0,p_h<0}^B (t^B)= t^B \left(1-\ell|p_h|\right) - \ell L \left( \alpha + \beta \right)|p_h| \ .
\end{gather}
This allows us to conclude that Bob, who is at the detector,
measures the following difference of  times of arrival between the soft photon
(detected at Bob at $t^B =0$) and the hard photon
\begin{equation}
\Delta t^B = \ell L |p_h| (\alpha + \beta) \ .
\label{delayFLAT}
\end{equation}

\begin{figure}[h!]
\includegraphics[width= 0.8\columnwidth]{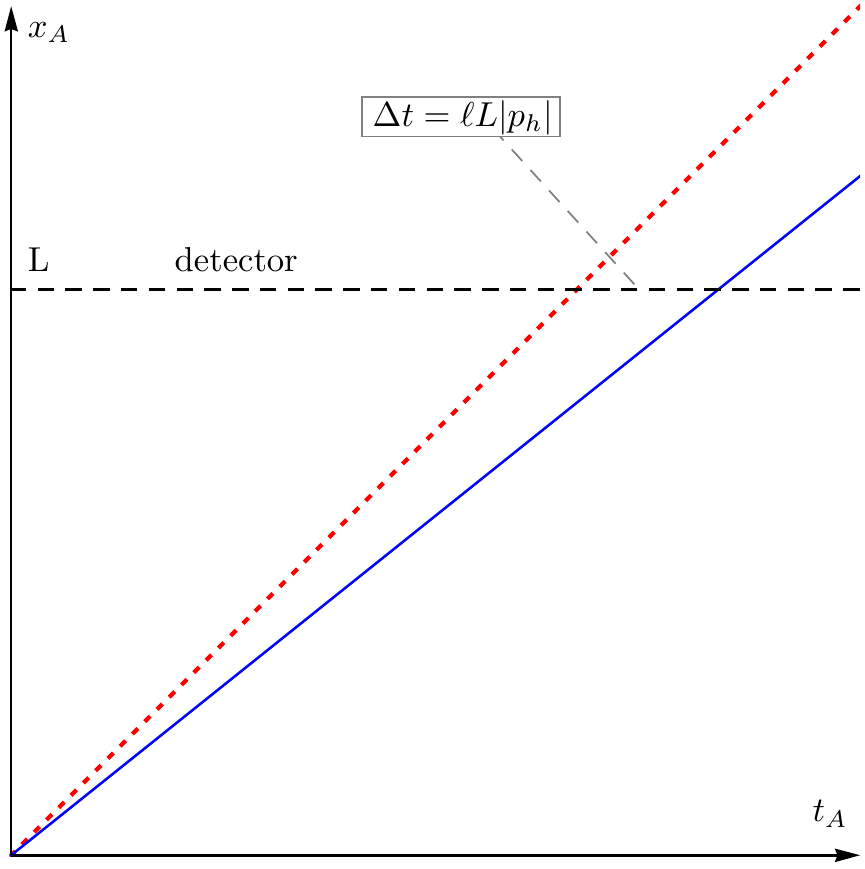}
\includegraphics[width= 0.8\columnwidth]{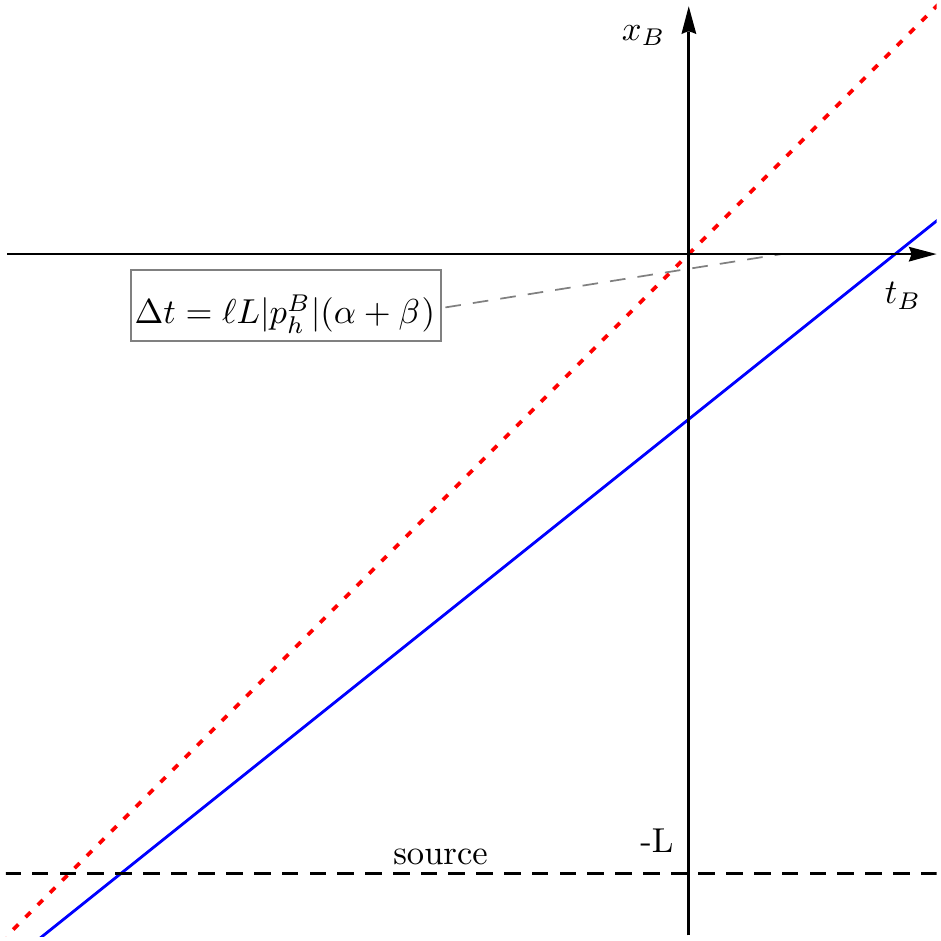}
\caption{We illustrate the results for travel times of massless particles derived in this section
by considering the case of two distant observers, Alice and Bob, in relative rest.
Two massless particles, one soft (dashed red) and one hard (solid blue), are emitted simultaneously at Alice and they reach Bob
at different times. Because of the nontriviality of translation transformations, in Bob's coordinatization
(bottom panel)
the emission of the particles at Alice appears not to be simultaneous. And similarly,
for the difference in times of
arrival at the distant detector (where Bob is located)
Alice finds in her coordinatization (top panel) a value which is not the same
as the difference in times of
arrival that Bob determines (bottom panel).
The case in figure has $\alpha + \beta =4/3$.
For visibility we assumed here unrealistic values for the scales involved.
For realistic values of the distance between observers and of the energy of the hard particle,
taking $\ell$ as the inverse of the Planck scale,
no effect would be visible in figure (the worldlines would coincide).}
\end{figure}

\newpage

So we see that the nontriviality of the translation transformations
does affect the difference between coordinate velocity and physical velocity.
And the relativity of locality produced by the nontriviality
of the translation transformations, while at first appearing to be counterintuitive,
actually ensures the internal logical consistency of the relativistic framework\footnote{As discussed
in greater detail in Refs.~\cite{whataboutbob,principle,kappabob}, there is an analogy between
the role of relative locality in deformed-Lorentz-invariant theories and the role
of relative simultaneity in deformed-Galilean invariant theories (ordinary Lorentz-invariant theories
are to be viewed, from this perspective, as deformations of Galilean-invariant theories,
which accommodate the invariant scale $c$ and the relativity of simultaneity, one being the
counterpart of the other).
The introduction of the invariant momentum scale $\ell^{-1}$ requires a deformation of Lorentz invariance
since ordinary Lorentz transformations change the value of all momentum scales. And
for the logical balance of the relativistic theory one finds that having one more thing
invariant (the scale $\ell$) requires rendering one more thing relative, which is the role played
by relative locality.
One can see that the same logics applies to the transition from Galilean Relativity to Einstein's Special Relativity.
Taking as starting point Galilean Relativity, the introduction of the invariant velocity scale $c$
requires a deformation of Galilean boosts,
since ordinary Galilean boosts change the value of all velocity scales. And
for the logical balance of the relativistic theory one finds that having one more thing
invariant (the scale $c$) requires rendering one more thing relative, which is the role played
by relative simultaneity.}.
Satisfactorily the physical velocity does depend on $\alpha$ and $\beta$ just in the way one
should expect on the basis of the role of $\alpha$ and $\beta$ in the on-shell relation.
But
the dependence of the physical velocity on $\alpha$ and $\beta$ is not very significant,
since it comes only in the combination $\alpha + \beta$.
It is because of this feature that
the difference between $E p^2$ deformation
and $E^3$ deformation, if analyzed in a flat/non-expanding spacetime, carries very little significance.
One can get the same physical velocity of massless particles by any mixture of $E p^2$ deformation
and $E^3$ deformation (including the cases where one of the two is absent)
as long as $\alpha + \beta$ keeps the same value.
There is nothing extraordinary or surprising about this: we are working at leading order in $\ell$,
so in terms which already have an $\ell$ factor we can use $E=|p|$ (for massless particles,
and using the fact that at zero-th order in $\ell$ the massless shell is $E=|p|$).
Evidently within these approximations here of interest a correction term of form $\ell E p^2$
is indistinguishable from a correction term of form $\ell E^3$.
While this is not surprising it was still worth devoting to it this section since one of the main
findings of the generalization proposed in the next sections is that essentially
in presence of spacetime expansion correction terms of form $\ell E p^2$
are very significantly different from corrections term of form $\ell E^3$.

Before moving on to those challenges, in closing this section let us comment briefly
on boost transformations. For the purposes we had in this section the relative locality
produced by deformed boost transformations did not come into play.
Readers interested in those features of relative locality can look in particular
at Ref.~\cite{whataboutbob}. The point relevant for us is that, as inferred from Ref.~\cite{whataboutbob},
the relative locality produced by boosts does not affect the derivation of the physical velocity,
but rather is the way by which the relativistic theory renders the properties of the physical
velocity compatible with boost invariance. This all comes down to the fact that $\ell$-deformed
boosts do not affect the timing of events at the observer: in the simple analysis reported above
an observer purely boosted with respect to Alice would also see as simultaneous the emission of the
two particles Alice sees as simultaneous; and another observer, this one purely boosted with respect
to Bob, would describe the timing of the detections of the two particles at Bob in a way
that is completely undeformed. Boosts
 do play a role in the analysis of coincidences of events distant from the
 observers\footnote{This point emerges most simply and clearly by analyzing, as done in Ref.~\cite{whataboutbob},
 the illustrative example of an observer Alice which describes two pairs of events, one pair coincident
 in her origin and one pair coincident far away from her origin.  For an observer purely boosted (purely $\ell$-deformed boosted) with respect to
 Alice one finds~\cite{whataboutbob} that the pair of events close to the origin are still coincident,
 but the pair of events distant from the origin are not coincident.
 This is the boost counterpart of the fact that, as a result of the properties of deformed
 translations, and the associated implications for relative locality,  the distant pair
 of events will not be observed as coincident by observers whose origin is close to that pair
 of events.},
 but play no role in the determination of the physical velocity.

\section{DSR-deformed de-Sitter-relativistic symmetries and physical velocity}

\subsection{DSR-deformed de-Sitter-relativistic symmetries
and equations of motion}

The preliminaries given in the previous section provide a good starting point
for the main challenge we intend to face in this manuscript. Those preliminaries
summarize some known aspects of $\ell$-deformed Lorentz symmetry for Minkowskian (non-expanding)
spacetimes, including the features of relativity of locality and the rather marginal relevance
of the differences between $E p^2$ deformations
and $E^3$ deformations. We now want to produce the first example of theory of classical-particle
worldlines with $\ell$-deformed de-Sitter-relativistic symmetries.
It is interesting that
de Sitter relativistic symmetries can themselves be viewed as
a deformation of the special-relativistic symmetries of Minkowski spacetime
such that the expansion-rate parameter $H$ is an invariant. And it is known that the invariance of the expansion rate
comes at the cost of some velocity artifacts.
With the points summarized in the previous section we can easily contrast the two
deformations of the special-relativistic symmetries of Minkowski spacetime which provide
the starting points for our work.
On one side we have $\ell$-deformed Lorentz(/Poincar\'e) symmetries, a deformation by a large
momentum scale $\ell^{-1}$
which produces velocity artifacts connected with the novel feature of relative locality.
And on the other side we have de-Sitter-relativistic symmetries,
a deformation by a large
distance scale $H^{-1}$
which produces velocity artifacts connected with spacetime expansion.
The theory we are seeking must be such that both of these features
can be accommodated, while preserving the
relativistic nature of the theory.
So both $\ell$ and $H$ shall be relativistic invariants (for a total of 3, including the speed-of-light scale,
here mute because of our choice of units). And we shall have both expansion-induced and relative-locality-induced
velocity artifacts.

The strength of these demands appears to confront us with an unsurmountable challenge. But
we shall see that
the ``preliminaries'' offered in the previous section draw a very clear path
toward our objective.
We start by specifying that our $\ell , H$-deformed relativistic symmetries shall leave invariant
the following combination
of the energy $E$, momentum $p$ and boost ${\cal N}$ charges of particles:
\begin{equation}
{\cal C}_{H, \alpha ,\beta}=E^{2}-p^{2}+2H{\cal N}p+\ell\left(\alpha E^{3}+\beta Ep^{2}\right) ~.
\label{casimir}
\end{equation}
Evidently for $\ell \rightarrow 0$ this reproduces the standard invariant of
de Sitter symmetries.
Something even more general than our correction term $\ell\left(\alpha E^{3}+\beta Ep^{2}\right)$
could here be envisaged, but we are not seeking results of maximum generality. On the contrary
we want to make the case as convincingly and simply as possible that
there are examples of the novel class of relativistic theories  we are here proposing.
Moreover the correction term $\ell\left(\alpha E^{3}+\beta Ep^{2}\right)$ does have enough structure
for us to investigate the interplay of $E p^2$ deformations
and $E^3$ deformations with spacetime expansion, which is the one among our side results that we
perceive as most intriguing.

The path drawn in the previous section guides us to observe that
the following $\ell$-deformed (2D) de Sitter algebra of charges is compatible
with the invariance of ${\cal C}_{H, \alpha ,\beta}$
\begin{gather}
\left\{ E,p\right\} =Hp-\ell\alpha HEp \ ,\nonumber \\
\left\{ E,{\cal N}\right\} =p-H{\cal N}-\ell\alpha E(p-H{\cal N})-\ell\beta Ep\ ,\nonumber \\
\left\{ p,{\cal N}\right\} =E+\frac{1}{2}\ell\alpha E^{2}+\frac{1}{2}\ell\beta p^{2}\ .
\end{gather}
One easily sees that for $\ell \rightarrow 0$ this reproduces the standard
properties~\cite{cacciatori}
of the classical de Sitter algebra of charges while for $H\rightarrow 0$ it reduces to (\ref{algebraFLAT}).

We give a convenient representation of these symmetry generators
in terms of ``conformal coordinates", with conformal time $\eta$
and spatial coordinate $x$,
and variables $\Omega,\Pi$ canonically conjugate to these conformal coordinates:
\begin{gather}
\left\{ \Omega,\eta\right\} =1\ ,\qquad\left\{ \Omega,x\right\} =0\ ,\nonumber \\
\left\{ \Pi,\eta\right\} =0\ ,\qquad\left\{ \Pi,x\right\} =1 ~,\nonumber \\
\left\{ \eta,x\right\} =0 ~.
\label{phasespaceCANONICAL}
\end{gather}
We find the following representation
\begin{gather}
\begin{split}
E & = -Hx\Pi+\Omega-H\eta\Omega +  \frac{\ell}{2}(1-\beta)\Pi^{2} \\&
~~~~+ \frac{\ell}{2}H\eta\Pi^{2} - \frac{\ell}{2}\alpha(\Omega -H\eta\Omega -Hx\Pi)^{2} \ ,
\end{split} \label{chargesTRANSLATIONSee} \\
p=\Pi\ ,~~~~~~~~~~~~~~~~~~~~~~~~~~~~~~~~~~~~~~~~~~~~~~~
\label{chargesTRANSLATIONSpp}
\end{gather}
\begin{equation}
\begin{split}
{\cal N}  = &\ x(1-H\eta)\Omega+\left(\eta-\frac{H}{2}\eta^{2}-\frac{H}{2}x^{2}\right)\Pi \\&
+\ell\left(\frac{1+H\eta}{2}x\Pi-(1-H\eta)\eta\Omega\right)\Pi\ .
\end{split}
\label{chargesBOOSTS}
\end{equation}

Again we intend to derive the worldlines of particles working within a ``covariant formulation";
so we formally introduce an auxiliary affine parameter on the worldline and the formal evolution
in the affine parameter is governed by the invariant ${\cal C}_{H, \alpha ,\beta}$.

Of course the fact that ${\cal C}_{H, \alpha ,\beta}$ is an invariant of the
(deformed-)relativistic symmetries implies that the charges that generate the symmetry transformations
are conserved over this evolution:
\begin{eqnarray}
&& \dot{E} \!=\! \{{\cal C}_{H, \alpha ,\beta},E\} \!=\! 0\ ,\ \dot{p} \!
=\! \{{\cal C}_{H, \alpha ,\beta},p\} \!=\! 0 \ , \nonumber\\
&& \qquad \qquad \dot{{\cal N}} \! =\! \{{\cal C}_{H, \alpha ,\beta},{\cal N}\} \!=\! 0\,. \nonumber
\end{eqnarray}
Importantly by consistency with the chosen (deformed) form of the invariant ${\cal C}_{H, \alpha ,\beta}$
we have obtained translation transformations which are significantly deformed, specifically for the time
direction. In fact, from (\ref{chargesTRANSLATIONSee}) and  (\ref{chargesTRANSLATIONSpp}) one finds
\begin{gather}
\left\{ E,\eta\right\} =1 \!-\! H\eta \!-\! \ell\alpha(1 \!-\! H\eta)(-Hx\Pi \!+\! \Omega \!-\! H\eta\Omega)\ ,\label{phasespaceET} \\
\begin{split}
\left\{ E,x\right\}  =  & - Hx +  \ell(\left(1 \!-\! \beta \!+\! H\eta \right)\Pi\\
& +\! \alpha H x ( \Omega \!-\! Hx\Pi \!-\! H\eta\Omega)),                                                                                                                                                         \end{split}
\label{phasespaceEX} \\
\left\{ p,\eta\right\} =0\ , \qquad \left\{ p,x\right\} =1 ~. \label{phasespaceC}
\end{gather}
This prepares us to deal with relative-locality effects, of the sort described in the previous section
but intertwined with the additional complexity of spacetime expansion.

But let us first note down the equations of motion that are obtained in our framework.
Following the standard procedures for derivation of worldlines in the covariant formulation
of classical mechanics one easily arrives,
for a particle of mass $m$ ({\it i.e.} ${\cal C}_{H, \alpha ,\beta} = m^2$),
at the following result for the worldlines:
\begin{equation}
\begin{split}
 x_{m,p} & = x_{0}+\frac{\sqrt{m^{2}+(1-H\eta)^{2}p^{2}}}{Hp} \\  &\,\,\,\,\, \,\,
-\frac{\sqrt{m^{2} +(-1+H\eta_{0})^{2}p^{2}}}{Hp} + \ell(\eta-\eta_{0})p\ .
\end{split}
\label{worldlineMASS}
\end{equation}
For the case of massless particles this reduces to
\begin{equation}
x_{m=0,p}\left(\eta\right)=x_{0}-\frac{p}{|p|}\left(\eta-\eta_{0}\right)\left(1-\ell|p|\right)\ ,
\end{equation}
which in turn specifying $p<0$ (so that the velocity is positive along the $x$-axis, see related comments in
the previous section) takes the form
\begin{equation}
x_{m=0,p}\left(\eta\right)=x_{0}+\left(\eta-\eta_{0}\right)\left(1-\ell|p|\right) ~.\label{worldline}
\end{equation}
By construction these
worldlines (\ref{worldlineMASS}),(\ref{worldline}) are covariant
under the (deformed-)relativistic transformations generated by the charges $E,p,{\cal N}$,
as one can also easily verify explicitly.

\subsection{Travel time of massless particles}
As already established in previous studies of the case without spacetime expansion
(here summarized in Section \ref{secflat}),
the DSR-deformed relativistic symmetries introduce (as most significant among
many other novel features) a dependence on energy of the travel time
of a massless particle from a given source to a given detector.
In the analysis we provided so far of our novel proposal
of DSR-deformed relativistic symmetries of an expanding spacetime (with constant expansion rate)
an indication of
dependence on energy of the travel time
of a massless particle from a given source to a given detector is found in Eq.~(\ref{worldline}):
the equation governing the properties of the worldline of a massless particle in conformal coordinates
is $\ell$-corrected and the correction term introduces a dependence on the
 energy(/momentum) of the particle.
However, as also expected on the basis of previous results on
the case without spacetime expansion, we are evidently working in a framework where locality is relative
(evidence of which has been provided so far within our novel framework implicitly
in Eqs.~(\ref{phasespaceET})-(\ref{phasespaceEX})).
So the analysis of the equations of motion written by one observer is inconclusive for what concerns
the notion of ``travel time", {\it i.e.} the correlation between emission time and detection time.
As illustrated for the non-expanding case in the previous section, we must guard against the coordinate
artifacts associated to relative locality by analyzing the emission of the particle in terms of
the description of an observer near that emission point and we must then analyze
the detection of the particle in terms of
the description of an observer near the detection point.

This is indeed our next task. We consider the case of two distant observers:
Alice at the emitter and Bob at the
detector. We keep the analysis in its simplest form, without true loss of generality,
by considering the case of simultaneous emission at Alice of only two massless particles,
one with ``soft" momentum $p_{s}$ and one with ``hard" momentum $p_{h}$.
Evidently, on the basis of the analysis reported in the previous subsection,
Alice describes the two particles according to
\begin{gather}
x_{p_{s}}^{A}(\eta^{A})=\eta^{A}\ ,\label{worldlinesALICEs} \\
x_{p_{h}}^{A}(\eta^{A})=\eta^{A}\left(1-\ell|p_{h}^{A}|\right)\ ,\label{worldlinesALICEh}
\end{gather}
where we specified $x_{0}^{A}=\eta_{0}^{A}=0$, so that the emission
is at $\left(0,0\right)^{A}$. Since translations are a relativistic symmetry
of our novel framework, we already know that the same two worldlines will be described
by the distant observer
Bob in the following way
\begin{gather}
x_{m=0,p_{s}}^{B}(\eta^{B})=x_{0;s}^{B}+\eta^{B}-\eta_{0;s}^{B}\ , \label{worldlinesBOBss} \\
x_{m=0,p_{h}}^{B}(\eta^{B})=x_{0;h}^{B}+\left(\eta^{B}-\eta_{0;h}^{B}\right)\left(1-\ell|p_h^{B}|\right)
 ~,\label{worldlinesBOBhh}
\end{gather}
{\it i.e.} the same type of worldlines but with a difference of parameters here codified
in $x_{0;s}^{B}$,$\,\eta_{0;s}^{B}$,$\,x_{0;h}^{B}$,$\,\eta_{0;h}^{B}$. Indeed,
Alice's worldlines (\ref{worldlinesALICEs})-(\ref{worldlinesALICEh})
and Bob's worldlines (\ref{worldlinesBOBss})-(\ref{worldlinesBOBhh})
have exactly the same form, but for the ones of Alice we had by construction
(by having specified simultaneous emission at Alice) that
$x_{0;h}^{A}=\eta_{0;h}^{A}=x_{0;s}^{A}=\eta_{0;s}^{A}=0$ whereas
Bob's values of the parameters, $x_{0;s}^{B}$,$\,\eta_{0;s}^{B}$,$\,x_{0;h}^{B}$,$\,\eta_{0;h}^{B}$,
should be determined by establishing which ($\ell$-deformed)
translation transformation connects Alice to Bob.

This is the same task performed in some of the previous studies
(such as Ref.~\cite{kappabob}) involving relative
locality in DSR-deformed relativistic theories without spacetime expansion.
But also this aspect of the analysis turns into a rather more challenging exercise for
us, having to deal with spacetime expansion. Let us then proceed
determining the translation transformation that connects Alice to Bob
providing
 enough details for the reader to appreciate the interplay between the different structures
 we must handle.

As usual we shall give the action on points $\eta,x$
of a worldline by
a finite transformation $T_{G;a}$ generated by the generic element $G$ of the relativistic-symmetry algebra
in terms
 of its exponential representation as
\begin{equation}
T_{G;a}
\triangleright X=e^{-aG}\triangleright X\equiv\sum_{n=0}^{\infty}\frac{\left(-a\right)^{n}}{n!}\left\{ G,X\right\} _{n}\,,
\label{finite_translation}
\end{equation}
 where $a$ is the transformation parameter and $\left\{ G,X\right\} _{n}$
is the $n$-nested Poisson bracket defined by the relation
\begin{equation}
\left\{ G,X\right\} _{n}=\left\{ G,\left\{ G,X\right\} _{n-1}\right\} \ ,\qquad\left\{ G,X\right\} _{0}=X\ .
\end{equation}

And before proceeding we must also implement something else that can be specified about observer Bob.
The worldline parameters $x_{0;h}^{A},\,\eta_{0;h}^{A},\,x_{0;s}^{A},\,\eta_{0;s}^{A}$
of observer Alice are fully specified ($x_{0;h}^{A}=\eta_{0;h}^{A}=x_{0;s}^{A}=\eta_{0;s}^{A}=0$)
by its being at the point of simultaneous emission of the two particles. We have introduced observer Bob
as one that detects the particles (the particles worldines should cross Bob spatial origin)
but we have so far left completely unspecified its
worldline parameters $x_{0;s}^{B}$,$\,\eta_{0;s}^{B}$,$\,x_{0;h}^{B}$,$\,\eta_{0;h}^{B}$.
We evidently can do better than that. And actually we can also insist, without true loss of generality,
that the soft particle reaches Bob in his spacetime origin:
so we are free to enforce $x_{0;s}^{B} = \eta_{0;s}^{B}=0$.

This is particulary convenient because it involves the ``soft particle", {\it i.e.} the one whose momentum $p_s$
has been chosen to be small enough to render the $\ell$-deformed effects inappreciable within the experimental
sensitivities available to Alice and Bob. The fact that both $x_{0;s}^{A}=\eta_{0;s}^{A}=0$
and $x_{0;s}^{B} = \eta_{0;s}^{B}=0$ for a soft particle
leads us to focus on the case of the observer Bob
connected to Alice by the following transformation
\begin{equation}
\mathrm{Bob} =e^{-a_{x}p}\triangleright e^{-a_{\eta}E}\triangleright \mathrm{Alice}\ ,\label{TTR}
\end{equation}
with
\begin{equation}
a_{x}=\frac{1-e^{-Ha_{\eta}}}{H} ~.
\label{parameterlink}
\end{equation}
Through this we are essentially exploiting the fact that the deformation is ineffective on
the soft particle as a way for us to focus on a distant observer Bob whose relationship to Alice
(translation parameters connecting Alice to Bob) can be specified using only known results
on the undeformed/standard
relativistic properties\footnote{This also implicitly requires~\cite{kappabob} that the clocks at Alice and Bob are synchronized
by exchanging soft massless particles.}.
Indeed in classical (relativistically undeformed) de Sitter spacetime one easily
finds that a massless particle emitted in the origin of some observer Alice will cross the origin
of all observers connected to Alice by  a conformal-time translation of parameter $a_\eta$
followed by a spatial translation of parameter $a_x$ with the request
that $a_{x}=H^{-1}[1-e^{-Ha_{\eta}}]$

Using our translation generators (\ref{chargesTRANSLATIONSee})-(\ref{chargesTRANSLATIONSpp})
one easily finds that for points on the worldline of the hard particle the map from Alice to Bob
is such that
\begin{gather}
\begin{split}
\!\!\!\!\!\!\!\!\!\!\!\!\!\!\!\!\!\!\!\!\!\!\!\!\!\!\!\!\!\!\!\!\!\!\!\!\!\!\!\!\!\!\!\!\!\!\!\!\! \!\!\eta_{h}^{B} \!=\! ~& \frac{1 \!-\! e^{Ha_{\eta}}}{H} \!+\! e^{Ha_{\eta}}\eta_{h}^{A}
\end{split}
\nonumber \\
\begin{split}
\,\,\,\,\,\,\,\,\,\,\,\,\,\,\,\,\,\,\,+ & \ell\,\alpha \, a_{\eta} e^{Ha_{\eta}} (1 - H\eta_{h}^{A}) \left( E_h^{A} + Ha_{x}p_{h}^{A}\right)\,,
\end{split}
\nonumber \\
\begin{split}
x_{h}^{B} =~ & e^{Ha_{\eta}} (x_{h}^{A} - a_{x} ) + \ell \beta \frac{\sinh\left(Ha_{\eta}\right) }{H} p_{h}^{A}\\
  -& \ell \alpha a_{\eta} e^{Ha_{\eta}} H(a_{x}-x_{h}^{A}) \left( E_{h}^{A} + Ha_{x} p_{h}^{A}\right) \\
-& \ell \frac{ \left(1-e^{-H a_\eta }\right) \left(1+e^{H a_\eta } H \eta_h^A \right) }{H}p_h^A\,,
\end{split}
\nonumber \\
p_{h}^{B}=e^{-Ha_{\eta}}\left(p_{h}^{A} + \ell\alpha Ha_{\eta} p_{h}^{A}\left(E_{h}^{A} + Ha_{x} p_{h}^{A}\right)\right)~.
\nonumber
\end{gather}
Crucial for us is the fact that, in light of this result for the laws of transformation from
Alice's $\eta_{h}^{A}, x_{h}^{A}, p_{h}^{A}$
to Bob's $\eta_{h}^{B}, x_{h}^{B}, p_{h}^{B}$,
we can deduce that the worldline of the hard particle emitted at Alice is described
by Bob as follows
\begin{equation}
x^B_{m=0,p_h}(\eta^{B})  \!=\! \eta^B \!\!- \ell |p_h^B| \!\left(\! \eta^B \!\!+\! \alpha a_\eta \!+\! \beta \frac{e^{2 H a_\eta } \!-\! 1 \!}{2 H} \right)\,,
\label{worldlineBOB}
\end{equation}
where we made use of all the specifications discussed above, including  $a_x = \frac{1-e^{-H a_\eta}}{H}$.

In turn this allows us to obtain the sought result for the
dependence on energy(/momentum) of the travel times of massless particles:
by construction of the worldlines and of the Alice$\rightarrow$Bob transformation the soft
massless particle emitted in Alice's spacetime origin reaches
 Bob's spacetime origin, whereas from (\ref{worldlineBOB}) we see that the hard massless particle also emitted in Alice's spacetime origin reaches Bob at a nonzero conformal time.
Specifically the difference in conformal travel times derivable from (\ref{worldlineBOB}) is
\begin{equation}
\begin{split}
\Delta \eta^B = \eta_h^B\Big|_{x_h^B=0} =\ell|p^{B}|\left(\alpha a_{\eta}+\beta\frac{e^{2Ha_{\eta}}-1}{2H}\right) ~.
\end{split}
\label{delayBOBold}
\end{equation}

We summarize the relativistic properties of this travel-time analysis, inconformal coordinates, in Fig.~2. By comparison with Fig.~1 one sees that  in conformal coordinates the  qualitative picture of the energy dependence
of travel times is very similar to the one of the case
\begin{figure}[h!]
\includegraphics[width= 0.8\columnwidth]{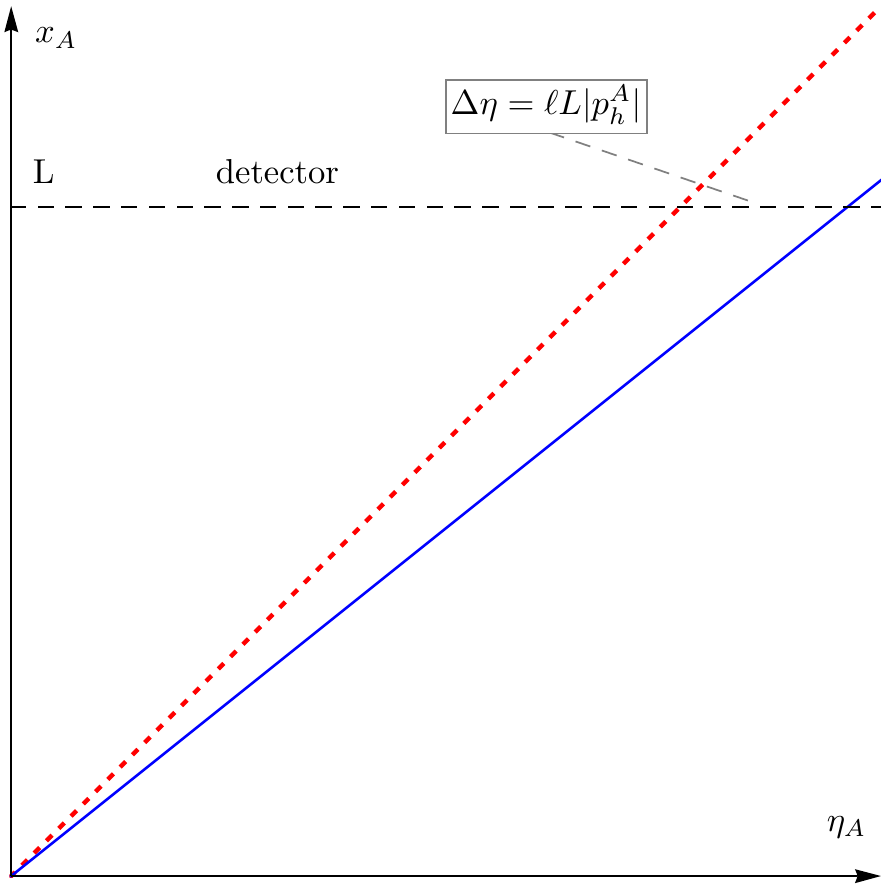}
\includegraphics[width= 0.8\columnwidth]{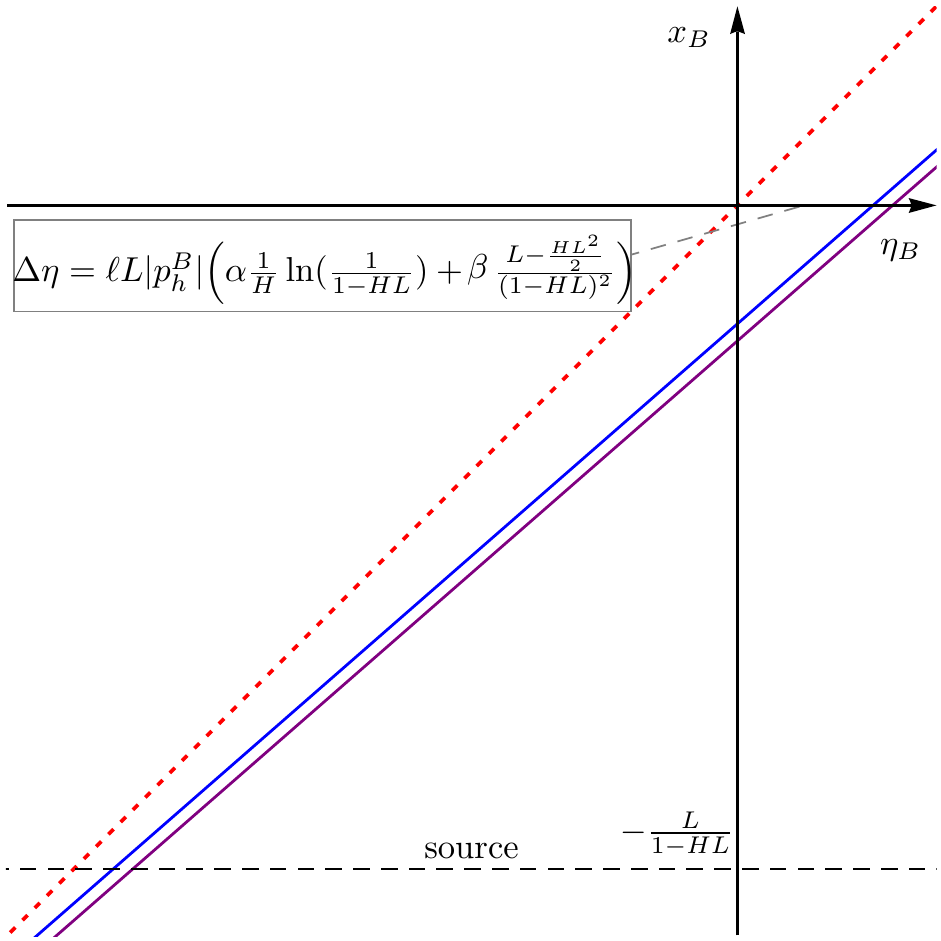}
\caption{We illustrate the results for travel times of massless particles derived in this section,
adopting conformal coordinates.
We consider the case of two distant observers, Alice and Bob, connected by a pure translation,
and two massless particles, one soft (dashed red) and one hard (solid blue and solid violet),
emitted simultaneously at Alice. And we consider two combinations of values of $\alpha$ and $\beta$
producing the same $\alpha + \beta$: the case $\alpha = 2/3, \beta = 2/3$
(Bob's hard worldline in blue)
and the case $\alpha = 1/3, \beta = 1$ (Bob's hard worldline in violet).
As in the case without expansion (Fig.~1) we have here that in Bob's coordinatization
(bottom panel)
the emission of the particles at Alice appears not to be simultaneous. Similarly for the difference in times of
arrival at the distant detector Bob
Alice finds in her coordinatization (top panel) not the same value as the difference in times of
arrival that Bob determines.
Comparison of the blue and violet worldlines shows that in the case with spacetime expansion
the travel time does depend individually on $\alpha$ and $\beta$ (not just on  $\alpha + \beta$
as in the case without spacetime expansion).
Again for visibility we assumed here unrealistic values for the scales involved.\vspace{-0.4cm}
}
\end{figure}

\noindent
without spacetime expansion. But here, with spacetime expansion, there are tangible differences between $E p^2$ deformations
and $E^3$ deformations.

In Fig.~3 we characterize the results of our travel-time analysis in comoving coordinates,
which for most studies of spacetime expansion are the most intuitive choice of coordinates.
[The differences between Fig.~2 and Fig.~3 all are a straightforward manifestation of the
relationship $\eta = H^{-1}(1-e^{-Ht})$
between the conformal time $\eta$ and the comoving time $t$.]

\begin{figure}[h!]
\includegraphics[width= 0.8\columnwidth]{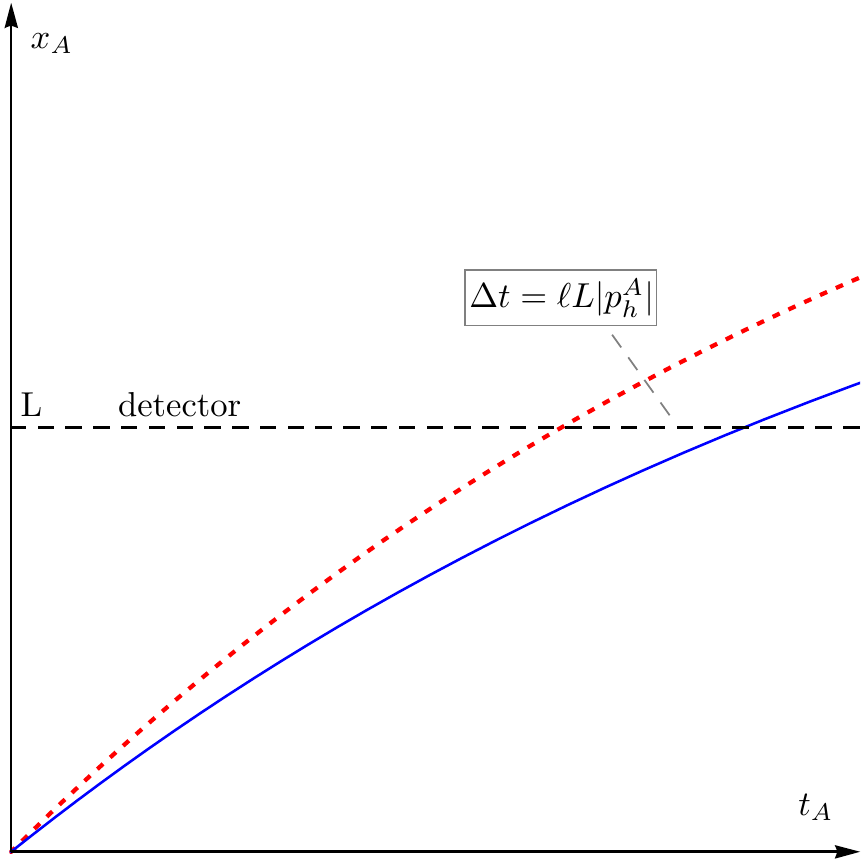}
\includegraphics[width= 0.8\columnwidth]{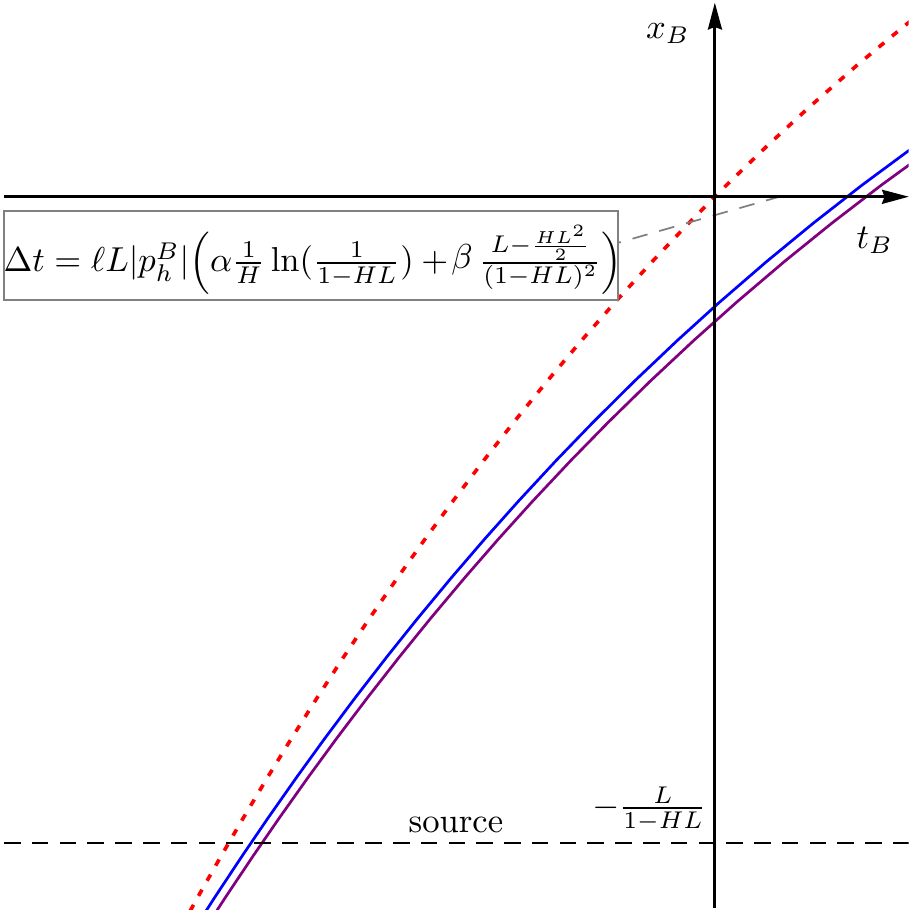}
\caption{We here illustrate the results for travel times of massless particles derived in this section,
adopting comoving coordinates. The situation here shown is the same situation already shown
(there in conformal coordinates) in Fig.~2.\vspace{-1cm}}
\end{figure}

\section{Implications for phenomenology}
Our main motivation for this analysis was conceptual. There is at this point rather robust evidence that
theories in non-expanding spacetime with DSR-deformed relativistic symmetries do have a solid internal logical
consistency and are therefore viable candidates for quantum-gravity researchers. Some of their properties,
like the relativity of spacetime locality,
are ``extremely novel" and this will understandably affect how
each researcher subjectively gauges the likelihood  (or lack thereof)
 of DSR-deformations of relativistic symmetries for the quantum-gravity realm.
But the objective fact is that, if we could confine our considerations to Minkowskian spacetimes,
the DSR option is at this point clearly viable.
However, it is of course a well established experimental fact that we are not in
a Minkowskian spacetime. Even postponing, as we did here, the much reacher collection of
experimental facts on gravitational phenomena and the Einsteinian description of spacetime, we felt
DSR research
 should at least start facing the fact that spacetime expansion is very tangible in our data.
 So we here sought to make the first step away from the
 extremely confining arena of Minkowskian spacetime,
 by showing that DSR-deformations of relativistic symmetries are also viable
 in deSitterian spacetime, expanding at a constant rate.

 In facing successfully this challenge, of primarily ``academic" interest,
   we stumbled upon results which are of rather sizable significance for phenomenology, for reasons
 that were hinted at in earlier parts of this manuscript and we shall now discuss in greater detail.

Our first step toward the phenomenological issues of interest in this section is to
reformulate the results for travel times of massless particles derived in the previous
section in a way that is in closer correspondence with the type of facts that are established
experimentally, when our telescopes observe distant sources of gamma rays.
Many of the analyses performed at our telescopes amount to timing the detection
of photons emitted from a source at a known redshift $z$. Postponing for a moment the fact that
the relevant contexts are not such that we could assume a constant expansion rate,
let us observe that the results we derived in the previous section are easily reformulated
as the following prediction for the differences in detection times
of photons of different energies(/momenta)
emitted simultaneously by a source at redshift $z$
\begin{equation}
\Delta t=\ell|p|\left(\alpha\frac{\ln\left(1+z\right)}{H}+\beta\frac{z+\frac{z^{2}}{2}}{H}\right)
~. \label{delayBOBz}
\end{equation}
Here again $\Delta t$ is the difference in detection times between a hard gamma-ray photon of momentum $p$
and a reference ultrasoft photon emitted simultaneously to the hard photon at the distant source. And notice
that with $\Delta t$ we denote the comoving time but in the relevant timing sequences
of detections at telescopes
the difference between
comoving  and conformal time is intangible: the telescopes are operated for a range of times within the $t=0$
of their resident clock which is relatively small, much smaller than the time scale $H^{-1}$,
so that conformal time and comoving time essentially coincide
(one has $\eta \equiv H^{-1}(1-e^{-Ht}) \simeq t$ for $t \ll H^{-1}$). For what concerns redshift
we simply relied on the standard formula \cite{jacktesi,DeSitter}
for the case of constant expansion rate which,
in our notation, gives $z=e^{Ha_{\eta}}-1$.

A first aspect of phenomenological relevance which must be noticed in our
result (\ref{delayBOBz}) is the dependence on the parameters $\alpha$ and $\beta$, {\it i.e.}
 the difference between $E p^2$ deformations
and $E^3$ deformations.
The fact that for the non-expanding spacetime case this dependence only goes with $\alpha + \beta$
is of course recovered in our result (\ref{delayBOBz}) in the limit of small redshift $z$.
This is expected since for small redshift the expansion does not manage to have appreciable consequences.
What is noteworthy is that already at next-to-leading order in redshift the dependence
on  $\alpha$ and $\beta$ is no longer fully of the form $\alpha + \beta$, as seen by
expanding our result (\ref{delayBOBz}) to second order in $z$:
\begin{equation}
\Delta t \Big|_{z \ll 1} \simeq
\frac{\ell|p|}{H}\left[(\alpha+\beta) z + (\beta - \alpha)\frac{z^{2}}{2}\right]
~. \label{delayBOBzAPPROX}
\end{equation}
In a non-expanding spacetime the difference between $E p^2$ deformations
and $E^3$ deformations would not have significant implications on travel-time determinations
(since it depends on $\alpha+\beta$ one gets the same result reducing the amount of $E p^2$ deformation in favor
of an equally sizable increase of the amount of $E^3$ deformation).
The situation in expanding spacetimes is evidently qualitatively different in this respect since
$E p^2$ deformations
and $E^3$ deformations produce corrections to the travel times that have different functional dependence on
redshift. So we are learning that for determinations of travel times from distant astrophysical sources
the difference between $E p^2$ deformations
and $E^3$ deformations is a phenomenologically viable (phenomenologically determinable) issue.

Related to this observation is also the other point we want to make on the phenomenology side,
which concerns the comparison with analogous studies of scenarios where relativistic symmetries
are actually ``broken" (allowing for a preferred/``aether" frame), rather than DSR-deformed as in the
case which was of interest for us in this manuscript. Broken-Lorentz-symmetry theories with a preferred frame
are far simpler conceptually than DSR-deformed relativistic theories, and indeed the issue
of the interplay between scale of Lorentz-symmetry breakdown and spacetime expansion has been usefully
investigated already for several years~\cite{ellisfirst,piranfirst,piransecond},
even producing a rather universal consensus on the
proper formalization that should be adopted~\cite{piransecond,ellisSECOND,grb090510}.
Again thanks to the
simplicity of broken-Lorentz-symmetry theories
these results apply to the general case of varying expansion rate, and for the  massless particles
they take the form
\begin{equation}
\Delta t = \lambda_{LIV} |p| \int_0^z \frac{dz}{a(z)H(z)} \ ,
\label{delayLIV}
\end{equation}
where $a$ is the scale factor, $H$ is the expansion rate,
and $\lambda_{LIV}$ (``LIV" standing for Lorentz Invariance
Violation) is the counterpart of our scale $\ell$: just like $\ell$ for us is the inverse-momentum scale
characteristic of the onset of the DSR-deformation of relativistic symmetries, $\lambda_{LIV}$ is
the inverse-momentum scale
characteristic of the onset of the LIV-breakdown of relativistic symmetries.
Of course, there is a crucial difference between the properties of $\ell$ and the properties of $\lambda_{LIV}$:
$\ell$, as characteristic scale of a deformed-symmetry picture, takes the same value for all observers,
while $\lambda_{LIV}$, as characteristic scale of a broken-symmetry picture, takes a certain value in the
preferred frame and different values in frames boosted with respect to the preferred frame.

Setting  momentarily these differences aside we can compare our results for DSR-deformed symmetries with
constant rate of expansion with
the special case of the broken-symmetry formula (\ref{delayLIV}) obtained
for constant rate of expansion:
\begin{equation}
\Delta t = \lambda_{LIV} |p| \frac{z + \frac{z^2}{2} }{H} \ .
\end{equation}
Comparing this to our result (\ref{delayBOBz}) we find that fixing $\alpha = 0$ in the deformed-symmetry case
one gets a formula (valid in all reference frames) which is the same as the formula of the
broken-symmetry case in the preferred frame. So if $\alpha =   0$ the differences between deformed-symmetry
and broken-symmetry cases would be tangible only by comparing studies of travel times of massless particles
between two telescopes with a relative boost: the difference there would be indeed that $\ell$ takes the same value
for studies conducted by the two telescopes whereas for $\lambda_{LIV}$  the two telescopes should give different
values.

We must stress however that within our deformed-symmetry analysis we found no reason to
focus specifically on the choice  $\alpha = 0$. And if  $\alpha \neq 0$ in the deformed symmetry case even studies
conducted by a single telescope could distinguish between the case of symmetry deformation and the case
of symmetry breakdown.
\vspace{-0.5cm}

\begin{figure}[h!]
\includegraphics[width= 1\columnwidth]{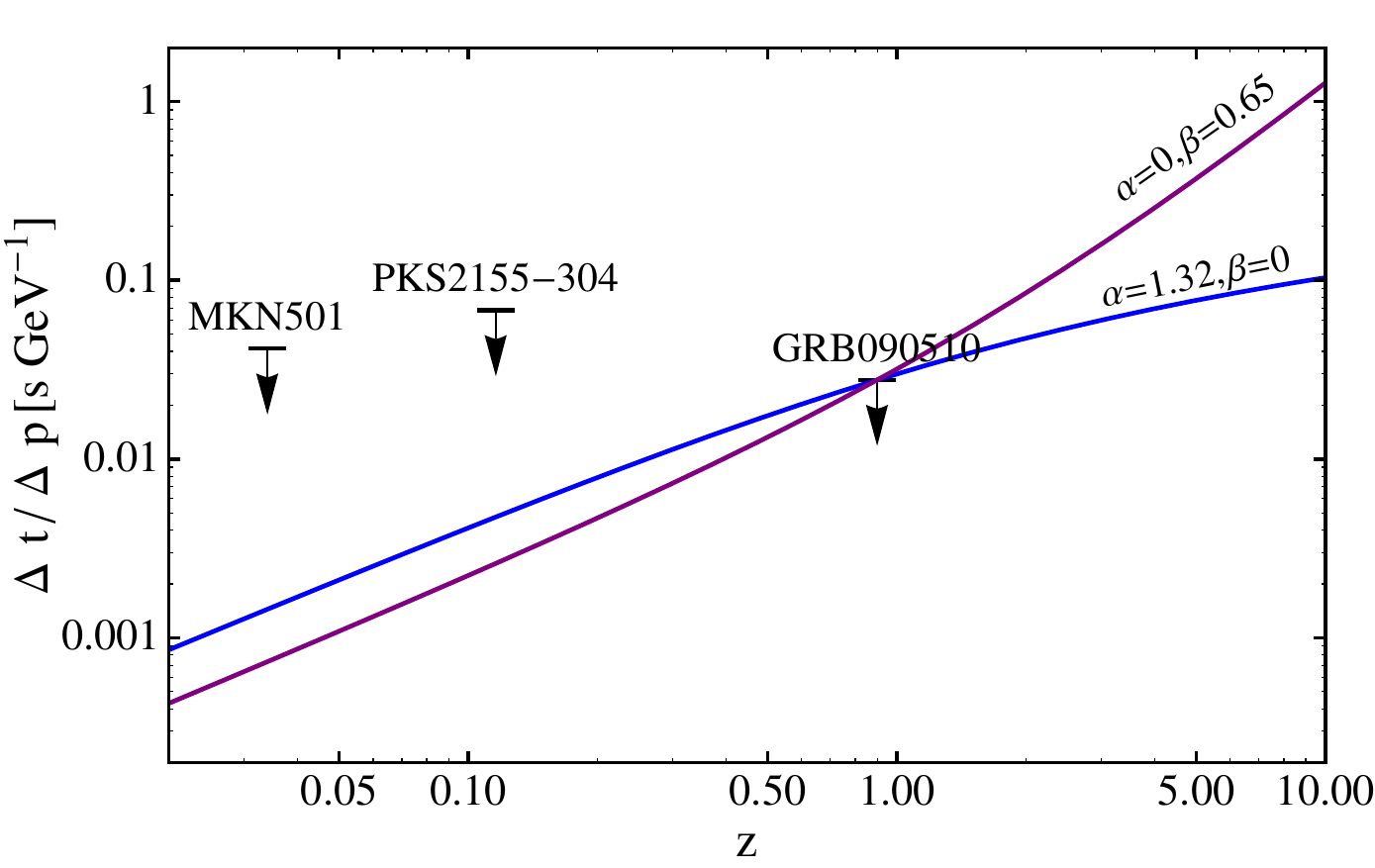}
\vspace{-0.7cm}\caption{Here we show the dependence on redshift of the expected time-of-arrival difference divided by the difference
of energy of the two massless particles. Two such functions are shown, one for the case $\alpha =0, \beta = 0.65$
 (violet) and one for the case $\alpha =1.3, \beta = 0$ (blue). We also show the upper limits that can be derived
 from data reported in Refs.~\cite{magic2007, hessVGAMMA, grb090510}, setting momentarily aside the fact that
 our analysis adopted the simplifying assumption
 of a constant rate of expansion (whereas a rigorous analysis of the data reported in
 Refs.~\cite{magic2007, hessVGAMMA, grb090510} should take into account the non-constancy of the expansion rate). The values $\alpha =0, \beta = 0.65$
 and  $\alpha =1.3, \beta = 0$ have been chosen so that we have consistency with the tightest upper bound, the one established in Ref.~\cite{grb090510}.  The main message is coded in the fact that at small values  of redshift the blue and the violet lines are rather close,  but at large values of redshift they are significantly different (this is a log-log plot). In turn this implies that at high redshift the difference between  adding correction terms of form $E p^2$ and adding correction terms of form $E^3$ can be very tangible.{\vspace{-1cm}}}
\end{figure}

\newpage

In Fig.~4 we compare the dependence on redshift of our deformed-symmetry effect among two limiting cases
of balance between $\alpha$ and $\beta$, and we also compare these results to
bounds on travel-time anomalies~\cite{magic2007, hessVGAMMA, grb090510}
obtained in studies of sources at redshift
smaller than 1
(where the assumption of a constant rate of expansion it not completely misleading).

In Fig.~5 we illustrate the ``constraining power" of our results and of foreseeable generalizations of our analysis gaining access to cases with non-constant rate of expansion.
For better visibility and as a way to offer more intelligible visual
messages we restrict our focus to the case in which both $\alpha$ and $\beta$ are positive. [There is no reason for making this assumption, but on the other hand it is easy to see how the constraints established within this assumption can be adapted to the more general situation with $\alpha$ and $\beta$ allowed
to also be negative.]

In the left panel of Fig.~5 we show how the bound obtained from Ref.~\cite{hessVGAMMA},
at the relatively small redshift
of $z=0.116$ (where we should really be able to apply our results for constant rate of expansion
as a reliable first approximation), provides a constraint on the $\alpha,\beta$ parameter space.
For these purposes we can of course fix the value of $\ell$ to be exactly the Planck
length (the inverse of the Planck scale) since any rescaling of $\ell$ can be reabsorbed
into an overall rescaling of $\alpha$ and $\beta$.
Within this choice of conventions the target ``Planck-scale sensitivity" would manifest
itself as the ability to constrain values of $\alpha$ and $\beta$ of order $1$.
As shown in the left panel of Fig.~5, even restricting our focus on cases with redshift much smaller
than $1$ (as needed because of the present limitation of applicability of
our approach to constant,
or approximately constant, rate of expansion) this Planck-scale sensitivity is not far.
In the right panel of Fig.~5 we show the much tighter constraint on the $\alpha,\beta$ parameter space
which would be within reach if we could assume our analysis to apply also to redshifts close to $1$,
as for GRB090510 observed by the Fermi telescope~\cite{grb090510}. Analyzing data from sources at redshift
of $\simeq 1$ assuming a picture with constant rate of expansion cannot produce conservative
experimental bounds, but the content of the right panel of Fig.~5 serves the purpose of
providing evidence of the fact that
full Planck-scale sensitivity will  be within reach of improved versions
of our analysis, extending our results to the case of expansion at non-constant rate.

\begin{figure}[h!]
\includegraphics[width= 0.49\columnwidth]{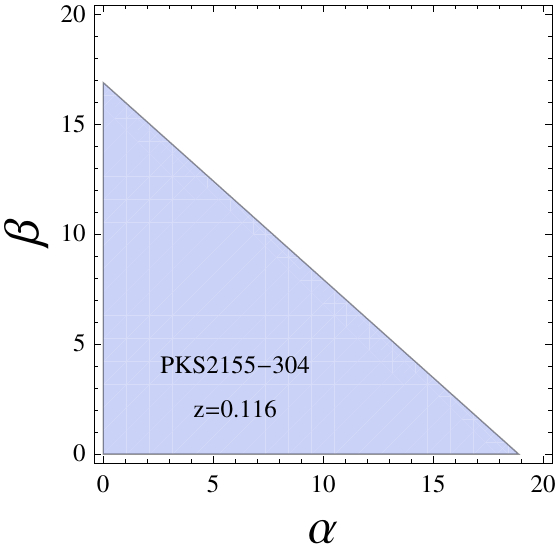}
\includegraphics[width= 0.49\columnwidth]{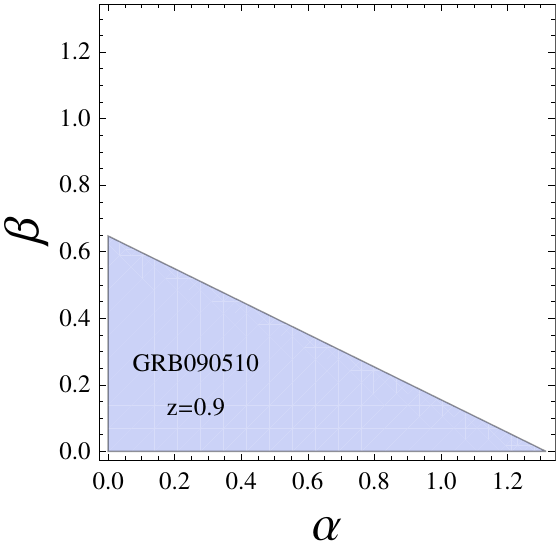}
\caption{Here in the left panel
 we show the constraint on the $\alpha,\beta$ parameter space
 that can be obtained from Ref.~\cite{hessVGAMMA}, concerning a source
at the relatively small redshift
of $z=0.116$, where we can confidently  apply our results for constant rate of expansion
as a reliable first approximation. Through this we show that even within the confines of our
analysis Planck-scale sensitivity
(values of $|\alpha|$ and $|\beta|$ smaller or comparable to $1$)
is not far.
In the right panel
 we show the much tighter (indeed ``Planckian") constraint on the $\alpha,\beta$ parameter space
which would be within our reach if we could assume our analysis to apply also to redshifts close to $1$,
as for GRB090510 observed by the Fermi telescope~\cite{grb090510}.
By comparing the left and the right panel one also finds additional evidence of how the
difference between  adding correction terms of form $E p^2$
and adding correction terms of form $E^3$ becomes more significant at higher redshifts:
in the left panel (data on source at small redshift of $z=0.116$) the bound on the $\alpha$ parameter is nearly as strong
as on the $\beta$ parameter, whereas in the right panel (data on source at
redshift of $z \simeq 0.9$) the constraint
on the alpha parameter is significantly weaker than the constraint on the $\beta$ parameter.}
\end{figure}

\section{Summary and outlook}
We here addressed a long-standing issue for the study of quantum-gravity-inspired deformations
of relativistic symmetries. It led us to propose
the first ever example of a relativistic theory of worldlines of particles
with 3 nontrivial relativistic invariants: a large speed scale (``speed-of-light scale"),
a large distance scale (inverse of the ``expansion-rate scale'),
and a large momentum scale (``Planck scale").
And on the basis of the observations reported in the previous section it is clear that
this is not merely an exercise providing a more rigorous derivation of previous
heuristic proposals: our constructive approach produces results that reshape the
targets of the relevant phenomenology.

In order to fully empower these phenomenological applications the next step would
be to generalize our approach in such a way to render it applicable to cases with
varying expansion rate.
For what concerns applications in astrophysics this will open the way to exploiting
data from the farthest sources, with the associated benefit of even larger distances
amplifying the minute Planck-scale effects.
And with the ability of handling varying expansion rates this research programme could also
gain access to data in cosmology. The main opportunity from this perspective comes from
the analyses of cosmology data which were at first inspired by varying-speed-of-light
theories (see, {\it e.g.}, Refs.~\cite{vsl1,alebrama,Machado:2011fq} and references therein).
In our deformed-relativistic theories the speed of light is not varying in the sense originally
intended: enforcing (however deformed) relativistic invariance we did not find room for
the speed-of-light scale to have different values between now and earlier epochs of the Universe,
but we did find a scenario for having that (with equal strength at all stages of the evolution
of the Universe) the speed of photons depends on their energy. This energy-dependence of the
speed of photons we found for scenarios with deformed relativistic
symmetries may affect cosmology in ways that are similar to
(and yet potentially interestingly different from~\cite{gacQDESITTER})
the varying-speed-of-light scenarios, simply because of the fact that the Universe was hotter
in its earlier stages of evolution, so that the typical energy of particles was higher.

Of course, besides possibly leaping toward these opportunities
available when generalizing our results to cases with non-constant expansion rate,
a lot more could be done with the constant-expansion-rate case.
We here managed, and only at the cost of facing some significant complexity,
to obtain a consistent theory of classical-particle worldlines, but of course
much more is needed before having established a consistent scheme
 of relativistic deformation of
our current theories. In this respect, a top
priority would be to formulate
quantum field theories with these deformed relativistic theories.

\section*{Acknowledgements}
We gratefully acknowledge conversations with Giulia Gubitosi, Salvatore Mignemi and Tsvi Piran. This work was supported in part by a grant from the John Templeton Foundation. AM also acknowledges support from a NSF CAREER grant.

\end{document}